\documentstyle[preprint,floats,pra,aps]{revtex}

\tightenlines  
\begin{document} \draft

\title{\LARGE \bf Quantum Tomography Approach in Signal Analysis}

\author{Margarita A. Man'ko}
\address{P.N. Lebedev Physical Institute, Leninskii Pr. 53,
Moscow 117924, Russia}

\maketitle

\begin{abstract}
Some properties of the fractional Fourier transform,
which is used in information processing, 
are presented
in connection with the tomography transform of optical signals.
Relation of the Green function of the quantum harmonic
oscillator to  the fractional Fourier transform is elucidated.
\end{abstract}

\vspace{8mm}

Analysis of signals (electromagnetic, acoustic, seismic, etc.,)
is based on studying the properties of a complex time-dependent 
function $f(t)$ (called ``analytic signal'') which describes  
a signal. Signal analysis is an essential ingredient of information 
processing.  The conventional method for studying a signal is  
Fourier analysis which provides a function 
$f_{\rm F}(\omega)$
describing the frequency structure of the signal. Fourier
analysis is equivalent to applying invertable map 
$f(t)\leftrightarrow f_{\rm F}(\omega)$
the analytic signal on
the Fourier component of the signal. Other methods 
to study signals, in which invertable maps of the analytic signal
function onto a function of two variables (time--frequency 
quasidistributions, for example, the Ville--Wigner 
quasidistribution~\cite{Wigner32,Ville48})
$f(t)\leftrightarrow W\left(t,\omega\right)$ 
are used, were introduced to describe a joint time--frequency 
distribution of the signal. 
These methods are intensively used in information processing.
If one makes the replacement
$f\rightarrow \Psi;\, t\rightarrow x,$
formally complex analytic
signal $f(t)$ is equivalent to the complex wave function
$\Psi(x)$ describing a system's state in quantum mechanics.
In view of this, results of quantum theory can be applied to
 signal analysis and vice versa. 

Recently, in quantum mechanics and quantum optics the invertable
tomography map of the wave function on the probability distribution
function of a random variable (depending also on  extra
parameters) was introduced. The application of this map to 
signal analysis (called ``noncommutative'' tomography of
analytic signal) was developed~\cite{Mendes}. 
 Advantages of the proposed tomorgaphic methods of  signal
analysis consist in the fact that they map a complex function 
(analytic signal) on the probability distribution which provides
the same information on a signal but elucidates the signal
properties in more visible appearance. 

Fourier transform of optical signals plays an important 
role in describing the shape and frequency content of optical
pulses (for the particular case of interferometric
methods of the investigation of output signals in semiconductor 
lasers, it was successfully applied in~\cite{IEEE,Poland}). 
Other transforms can be used for an analysis of optical
signals for describing both their amplitude and phase; thus, 
the fractional
Fourier transform~\cite{Namias} 
was intensively employed in optical
measurements and information processing~\cite{FT5,FT1}.
In quantum optics, the symplectic tomography transform was
introduced~\cite{Tombesi95} 
to describe a quantum state, which as well
can  be conventionally described by the Wigner quasidistribution 
function~\cite{Wigner32}. 
This transform is an extension of 
Fourier transform which was also used in the optical tomography 
procedure~\cite{Vogel89,Raymer92} 
to describe a quantum state. 
The cited transforms can
be determined by a kernel of the integral operator. Analogously,
in quantum mechanics the Green function, for example, 
of the quantum harmonic oscillator
is the kernel of the quantum time-evolution operator. 
The aim of this contribution is to discuss the similarity of the Green 
function of the harmonic oscillator and the kernel of the fractional
Fourier transform. We also establish a connection of  the fractional
Fourier transform to the tomography method suggested
for measuring quantum states~\cite{Tombesi95}.

 The wave function of the 
harmonic oscillator 
$\Psi\left(x,t\right)$ satisfies the Schr\"odinger evolution equation
(in the coordinate representation)
\begin{equation}\label{fr2}
i\hbar \,\frac{\partial\Psi\left(x,t\right)}{\partial t}=
-\, \frac{\hbar^2}{2m}\frac{\partial^2\Psi\left(x,t\right)}{\partial x^2}
+\frac {m\omega^2x^2}{2}\, \Psi\left(x,t\right).
\end{equation}
The Green function $G_\omega \left(x,y,t\right)$ 
of the Schr\"odinger equation~(\ref{fr2}) determines the wave function
$\Psi\left(x,t\right)$ in terms of the initial wave function
$\Psi\left(y,0\right)$ 
by the relationship
\begin{equation}\label{fr3}
\Psi\left(x,t\right)=\int G_\omega \left(x,y,t\right)
\Psi\left(y,0\right)\,dy\,.
\end{equation}
The initial value of the Green function is
$G_\omega \left(x,y,0\right)=\delta \left(x-y\right).$
The Green function of the harmonic oscillator reads
\begin{equation}
\label{fr17}
G_\omega \left(x,y,t\right) =\sqrt {\frac {m\omega}
{2\pi i\hbar \sin \omega t}}\,\exp \left\{\frac {im\omega}{2\hbar}
\left[\left(x^2+y^2\right)\mbox{cot}\,\omega t
-\frac {2xy}{\sin \omega t}
\right]\right\}.
\end{equation}
Inserting (\ref{fr17}) in (\ref{fr3}) provides the explicit form of 
the integral 
transform of the initial wave function $\Psi\left(y,0\right)$
\begin{equation}
\label{fr18}
\Psi\left(x,t\right) =\sqrt {\frac {m\omega}
{2\pi i\hbar \sin \omega t}}\,\int \Psi\left(y,0\right)
\exp \left\{\frac {im\omega}{2\hbar}
\left[\left(x^2+y^2\right)\mbox{cot}\,\omega t
-\frac {2xy}{\sin \omega t}
\right]\right\}~dy\,.
\end{equation}
One can see that for time $t$ satisfying the condition 
$\mbox {cot}\, \omega t=0,$
the integral transform~(\ref{fr18}) coincides with the usual Fourier 
transform of the initial wave function, i.e.,
for $\omega t=\left({\pi}/{2}\right)+2\pi k,\,k=0,\pm 1,\pm 2,\ldots ,$
one has $\sin \omega t=1,$ and Eq.~(\ref{fr18}) reads 
\begin{equation}
\label{fr19}
\Psi\left(x, t=\frac {1}{\omega}\left[\frac {\pi}{2}+2\pi k\right]
\right) =\sqrt {\frac {m\omega}
{2\pi i\hbar }}\,\int \Psi\left(y,0\right)
\exp \left\{-\frac {im\omega}{\hbar}\,xy\right\}~dy\,.
\end{equation}
For arbitrary time $t$, 
relation~(\ref{fr18}) is the integral 
transform of the initial wave function with the 
Gaussian kernel, periodic in time. 

In  signal analysis and information processing, 
the fractional Fourier transform  
$\left({\cal F}^aq\right)\left(u\right)$
of  analytic signal 
$q\left(u\right)$
is used (see, for example,~\cite{FT5})
\begin{equation}\label{F1}
\left({\cal F}^aq\right)\left(u\right)=
\int B_a\left(u,u'\right)q\left(u'\right)~du'.
\end{equation}
The kernel of the transform reads
\begin{equation}\label{F2}
B_a\left(u,u'\right)=\exp \left[-i\left(\frac{\pi\hat\Phi}{4}
-\frac {\Phi}{2}\right)\right]|\sin\Phi|^{-1/2}
\exp\left[i\pi\left(u^2\cot \Phi-
\frac {2uu'}{\sin\Phi}+u'^2\cot\Phi\right)\right],
\end{equation}
where the angle variable is determined by the real parameter $a$,
$\Phi=a\pi/2,$ $0<|a|<2,$ and $\hat\Phi=\mbox{sgn}
\left(\sin\Phi\right).$
For $a=0$ and $a=2$, 
$B_0\left(u,u'\right)=\delta\left(u-u'\right)$
and $B_2\left(u,u'\right)=\delta\left(u+u'\right),$
respectively. 
The fractional Fourier transform is the linear transform.
For $a=1$ (the first-order transform), it
corresponds to the usual Fourier transform.

Let us now compare relations~(\ref{F1}), (\ref{F2}), 
and (\ref{fr18}) using the change 
of  variables
$$u=\frac {x}{\sqrt{2\pi}}\,\sqrt{\frac {m\omega}{\hbar}}\,,\qquad
u'=\frac {y}{\sqrt {2\pi}}\,\sqrt{\frac {m\omega}{\hbar}}\,,\qquad
\omega t=\Phi\,,$$
and  the replacement
$$
\left(\frac {\hbar}{m\omega}\right)^{1/4}
\Psi\left(\sqrt{\frac{2\pi \hbar}{m\omega}}
\,u,\,\frac {\Phi}{\omega}\right)\Longrightarrow
\left({\cal F}^aq\right)\left(u\right),\qquad
\left(\frac{\hbar}{m\omega}\right)^{1/4}
\Psi\left(\sqrt{\frac{2\pi\hbar}{m\omega}}
u,\,0\right)\Longrightarrow q\left(u\right).
$$
We see that (\ref{F1}), (\ref{F2}), 
and (\ref{fr18}) coincide up to the factor 
$\exp\left(i\Phi/2\right)
=\exp\left(i\omega t/2\right),$
this means that the identity of the oscillator's Green function 
and the kernel of the fractional Fourier transform takes plase,
i.e.,
$$
B_a\left(u,u'\right)
=\exp\left(\frac {i\omega t}{m\omega}\right)
G_\omega\left(x,y,t\right),
\quad x=\sqrt{\frac{2\pi\hbar}{m\omega}}\,u\,,
\quad
y=\sqrt{\frac{2\pi \hbar}{m\omega}}\,u',
\quad
t=\frac{\pi a}{2\omega}\,.
$$
The other phase factor in the kernel of the 
fractional Fourier transform is equal to the constant
phase factor of the Green function
$\exp\left(-i\pi \hat \Phi/4\right)
=i^{-1/2}.$

In~\cite{Mendes}, 
the procedure of  
noncommutative tomography of the analytic 
signal was suggested which  uses the symplectic 
tomography approach  for measuring
quantum states in quantum mechanics 
proposed in~\cite{Tombesi95}. 
Below, we
show the relation of the fractional Fourier transform to the 
noncommutative tomography approach. 
The marginal probability distribution 
$w\left(X,\mu,\nu\right)$ is connected 
to analytic signal $q\left(u\right)$ by means of the 
relationship~\cite{Mendes}
\begin{equation}\label{56}
w\left(X,\mu,\nu\right)
=\frac {1}{2\,\pi|\nu|}\left|\int q\left(u\right)
\exp\left(\frac {i\mu}{2\nu}\,u^2
-\frac {iX}{\nu}\,u\right)~du\right|^2.
\end{equation}
For arbitrary real parameters $\mu$ and $\nu,$
the probability distribution is normalized
$\int w\left(X,\mu,\nu\right)~dX=1,$
if  analytic signal is normalized
$\int|q\left(u\right)|^2~du=1.$
If one knows 
$w\left(X,\mu,\nu\right)$,  analytic signal 
can be reconstructed, in view of the relationship
\begin{eqnarray}\label{59}
q\left(u\right)\,q^*\left(u'\right)
=\frac {1}{2\,\pi}\int w\left(X,\mu,u-u'\right)
\exp\left[i\left(X-\mu\,\frac {u+u'}{2}\right)\right]~dX~d\mu\,.
\end{eqnarray}
After inserting (\ref{56}) in (\ref{59}), one obtains that 
the product 
of the analytic signal functions 
\begin{eqnarray}\label{60}
q\left(u\right)\,q^*\left(u'\right)
&=&\frac {1}{\left(2\,\pi\right)^2}\int\exp\left[i\left(X-\mu\,
\frac {u+u'}{2}\right)\right]
\left[\mu^2+\left(u-u'\right)^2\right]^{-1/2}~dX~d\mu
\nonumber\\
&&\times\,\frac {1}{\left|\sin\,\mbox{arctan}\,
\left[\left(u-u'\right)/\mu\right]\right|}
\left|\int \exp\left\{\frac {i}{2}
\left[\mbox{cot}\,\left(\mbox{arctan}\,
\frac {u-u'}{\mu}\right)y^2\right.\right.\right.\nonumber\\
&&\left.\left.\left.
-\,\frac {2X}{\sqrt{\mu^2+\left(u-u'\right)^2}}\,\frac{y}{\sin
\left[\mbox{arctan}\,\left(u-u'\right)/\mu\right]}\right]
\right\}q(y)~dy\right|^2
\end{eqnarray}
is expressed in terms of modulus of the fractional Fourier 
transform. 

In fact, let us compare (\ref{F2}) 
where the arguments $u,u'$ of the kernel $B$ are replaced with
$v,v'$ and the two last lines in formula~(\ref{60}). One 
can see that the change of the variables  
$\Phi \rightarrow \mbox{arctan}\,\left[\left(u-u'\right)/\mu\right]$ 
in (\ref{F2})
and $y\rightarrow \sqrt{2\pi}v';$ $X\rightarrow 
v\,\sqrt{\mu^2+\left(u-u'\right)^2}$ in (\ref{60})
gives to formula~(\ref{60}) the form similar to (\ref{F2}), namely,
$$\left|\int B_a\left(v,v'\right)
q\left(v'\right)~dv'\right|^2,\qquad \mbox{with}\qquad
 a=\left(2/\pi\right)\mbox{arctan}\,
\left[\left( {u-u'}\right)/{\mu}\right].$$

Thus, the connection between the noncommutative tomography approach
of~\cite{Mendes} 
and employment of the fractional 
Fourier transform is established.
The important aspect of applying the 
fractional Fourier transform in this
context is that in order to reconstruct 
 analytic signal (up to constant phase)
one needs only 
a modulus of the transform and this modulus 
has the meaning of the 
probability distribution function 
depending on  two real parameters.

In conclusion, we would like to point out that we
have demonstrated  the formal similiarity (better to say
even identity) of the fractional Fourier transform used 
in information processing and signal analysis
and the time-evolution transform of the wave function
of the quantum harmonic oscillator. The kernel of the fractional 
Fourier transform is mathematically  equivalent to the Green 
function of the quantum harmonic oscillator.  
This observation gives the possibility
to use physically obvious properties of the Green function like 
unitarity of the evolution operator to describe the properties 
of the fractional Fourier transform.
The experimentai realization of the fractional Fourier transform
can be done in optical fibers (selfoc) where the signal propagation
is described by a Schr\"odinger-like equation in the Fock--Leontovich
approximation~\cite{Gevork,Roma}. 

We have shown that the fractional Fourier transform is connected 
with the symplectic tomography approach of measuring quantum 
states and with the 
noncommutative tomography of analytic signals. 
The observed relations of
 quantum problems to some procedures used to analyze different
(for example, optical) signals provide the idea to use  
Green functions of
quantum systems with other potentials as kernels of transforms 
of analytic
signals, the kernels being  
different from the kernel of the fractional Fourier transform 
related to the harmonic oscillator potential. 
All these Green functions have the property of unitarity and 
by inverting time one has the 
kernel of inverse transform related to the Green functions.

Detailed description of the results of this study is done in~\cite{jrlr}.

\end{document}